\documentclass[]{spie}  

\usepackage{amsmath,amsfonts,amssymb}
\usepackage{subcaption}
\usepackage{graphicx}
\usepackage{cite} 
\usepackage{epsfig}
\usepackage{amsmath}
\usepackage{nccmath}
\usepackage{amssymb}
\usepackage{mwe}
\usepackage{acro}
\usepackage{amssymb}
\usepackage{xcolor,colortbl}
\usepackage{tabularx}
\usepackage{relsize}
\usepackage{pifont}
\usepackage{booktabs} 
\usepackage{multirow}
\usepackage{multicol}
\usepackage{adjustbox}
\usepackage{float}
\usepackage{graphicx}
\usepackage{makecell}
\usepackage{tabu}
\usepackage[colorlinks=true, allcolors=blue]{hyperref}
\usepackage[capitalize]{cleveref}

\usepackage{booktabs}  

\title{Cross-Species Data Integration for Enhanced Layer Segmentation in Kidney Pathology}
   
\author[a]{Junchao Zhu}
\author[b]{Mengmeng Yin}
\author[a]{Ruining Deng}
\author[d]{Yitian Long}
\author[b]{Yu Wang}
\author[c]{Yaohong Wang}
\author[c]{Shilin Zhao}
\author[b]{Haichun Yang}
\author[a,b,d]{Yuankai Huo*}
\affil[a]{Department of Computer Science, Vanderbilt University, Nashville, TN, USA}
\affil[b]{Department of Pathology, Microbiology and Immunology, Vanderbilt University Medical Center, Nashville, TN, USA}
\affil[c]{Department of Biostatistics, Vanderbilt University Medical Center, Nashville, TN, USA}
\affil[d]{Data Science Institute, Vanderbilt University, Nashville, TN, USA}

\authorinfo{Corresponding author: Yuankai Huo: E-mail: yuankai.huo@vanderbilt.edu}

\begin{document} 
\maketitle
\begin{abstract}
Accurate delineation of the boundaries between the renal cortex and medulla is crucial for subsequent functional structural analysis and disease diagnosis. Training high-quality deep-learning models for layer segmentation relies on the availability of large amounts of annotated data. However, due to the patient's privacy of medical data and scarce clinical cases, constructing pathological datasets from clinical sources is relatively difficult and expensive. Moreover, using external natural image datasets introduces noise during the domain generalization process. Cross-species homologous data, such as mouse kidney data, which exhibits high structural and feature similarity to human kidneys, has the potential to enhance model performance on human datasets. In this study, we incorporated the collected private Periodic Acid-Schiff (PAS) stained mouse kidney dataset into the human kidney dataset for joint training. The results showed that after introducing cross-species homologous data, the semantic segmentation models based on CNN and Transformer architectures achieved an average increase of 1.77\% and 1.24\% in mIoU, and 1.76\% and 0.89\% in Dice score for the human renal cortex and medulla datasets, respectively. This approach is also capable of enhancing the model's generalization ability. This indicates that cross-species homologous data, as a low-noise trainable data source, can help improve model performance under conditions of limited clinical samples. Code is available at \url{https://github.com/hrlblab/layer_segmentation}.

\end{abstract}


\section{INTRODUCTION}
\label{sec:intro}  

A kidney biopsy can provide important information about a patient's kidney function and is a crucial standard for diagnosing kidney-related diseases\cite{hogan2016native}. During histopathological analysis of kidney biopsy tissues, precise localization of the boundary between the cortex and medulla layers is essential for subsequent functional structural analysis (such as the glomerulus) \cite{stevens2006assessing, rosenberg2017focal}. Moreover, the renal cortex and medulla play distinct roles in maintaining bodily homeostasis\cite{dumas2021phenotypic}. Changes in the boundaries and proportions of the renal cortex and medulla are associated with the occurrence and progression of diseases such as chronic kidney disease and renal tumors\cite{windpessl2023preventing, wolf2018magnetic}. In clinical practice, this process is typically performed by experienced doctors through visual inspection under a microscope. However, due to the large size of histopathological images, manually observing and localizing the layers of kidney tissue under a microscope is time-consuming and labor-intensive, which motivates people to develop efficient automated tools.

In recent years, data-driven deep-learning methods\cite{he2016deep, ronneberger2015u,ke2023clusterseg} have introduced new tools for clinical computer-aided diagnosis, significantly improving diagnostic efficiency and accuracy. The popularity of convolutional neural networks (CNNs) and Transformer architectures has contributed to a variety of clinical practices\cite{zhu2023anti,ke2024tshfna}, including the segmentation of retinal fundus images \cite{ting2017development} and the diagnosis of breast cancer\cite{adam2023deep}. In the field of kidney diseases, deep-learning models have been employed to identify and segment glomerular structures in human kidney biopsies. However, training high-quality neural network models demands a substantial amount of high-quality annotated data. In the medical field, data collection from humans is particularly challenging due to medical ethics reviews and patient privacy concerns. 

To address this issue, seeking relevant external data has become a potential approach. Some studies have improved the performance of medical image classification by using transfer learning with models pre-trained on natural images (e.g., ImageNet)\cite{kim2022transfer}. However, due to the unique staining styles and tissue structures of histopathology images, noise may be introduced during domain generalization. In contrast, homologous organs across species have functional and structural similarities. Some studies have shown a high degree of similarity between human and mouse kidney tissues \cite{lindstrom2018conserved}, with common features such as localized lesions and inflammatory tissues found in pathological conditions like kidney tumors and diabetic nephropathy in both species \cite{burhans2018contribution, sato2020developmental}.

\begin{figure*}
\begin{center}
\includegraphics[width=\linewidth]{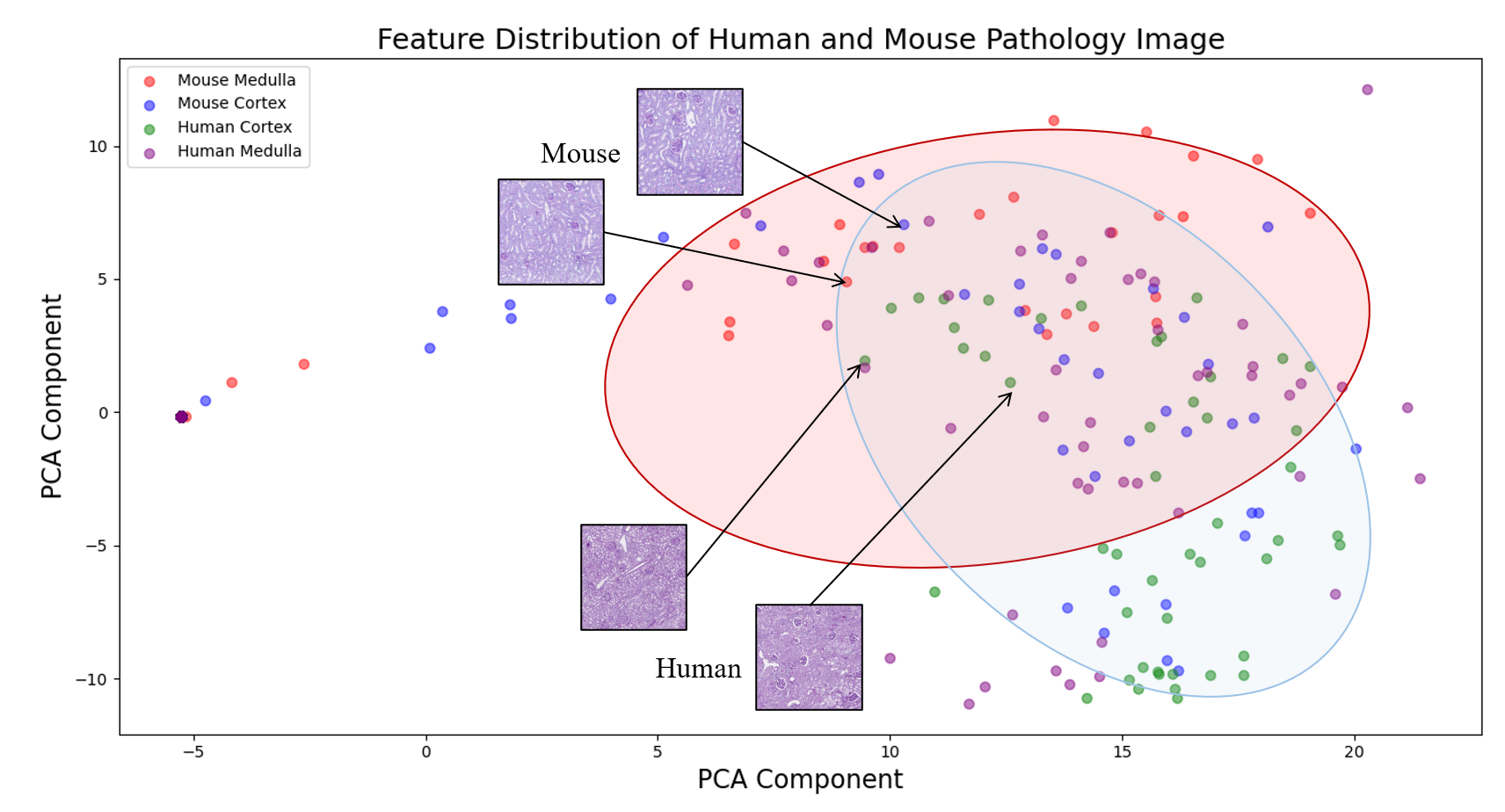}
\end{center}
\caption{The feature distribution of the cortex and medulla in human and mouse kidney images at the patch level. Human and mouse data points have a high overlap in the PCA space, indicating a high degree of cross-species similarity in the homologous tissue structure.}
\label{fig:similarity}
\end{figure*}

The feature distribution of the cortex and medulla in human and mouse kidney images at the patch level is shown in Figure. \ref{fig:similarity}. We extracted features from 50 histopathological images of human and mouse kidneys by ResNet50 and then reduced the dimension of features by performing the principal component analysis (PCA). The figure shows a high degree of overlap between the data points of humans and mice in the PCA space, indicating a high cross-species similarity in these two image regions. Therefore, data from homologous organs across species may improve model performance and generalization ability on human datasets.

In this study, we demonstrated that incorporating cross-species homologous data for collaborative training can enhance the performance of models in segmenting the cortex and medulla in human kidney histopathology images. We combined a private mouse kidney dataset stained with Periodic Acid-Schiff (PAS) with human kidney datasets for joint training. Compared with training on each dataset separately, the cross-species mixed data training achieved performance improvements across all targets in mainstream CNN and Transformer-based segmentation networks. This approach allows the model to leverage knowledge from the mouse dataset and adapt it to the human dataset, thereby improving segmentation performance and generalization capabilities

\section{METHOD}
\label{sec:method}  
\subsection{Cross-species Training Framework}

Before training the model, we first annotated the boundaries between the cortex and medulla at the whole slide image (WSI) level in the human and mouse datasets, completing the annotations for semantic segmentation. Subsequently, we cropped the WSIs into 1024×1024 pixel patches. These patches are then used to train and test several baseline models, including separate treatments for each species' data as well as combined training processing.

\begin{figure*}
\begin{center}
\includegraphics[width=\linewidth]{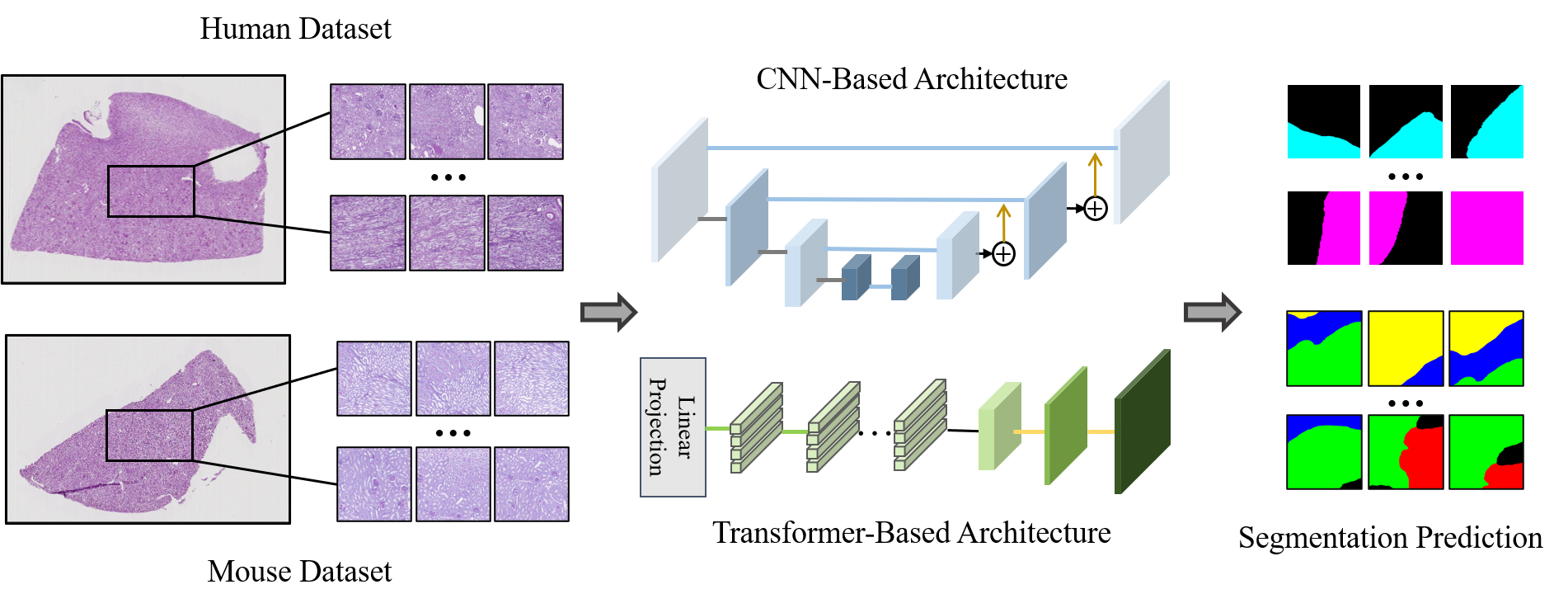}
\end{center}
\caption{Framework of the Cross-Species Training Process: Patches are first extracted from WSIs and then are utilized to train and test across several baseline models, which includes individual treatments for each species' data as well as combined data processing.}
\label{fig:Framework}
\end{figure*}

The training process is split into two distinct parts when training independently, including a four-class segmentation task for the mouse kidney dataset and single-category segmentation tasks for two categories within the human kidney dataset. To handle this diversity, we employ a hybrid loss function that combines Cross Entropy Loss (CE) and Dice Loss. Cross Entropy Loss evaluates the classification accuracy of each pixel in multi-class segmentation tasks, while Dice Loss quantifies the overall similarity between the predictions $y'$ and the ground truth $y$. Thus, the hybrid loss function $L_s$ in independent training can be stated as:
\begin{equation}
L_s=\lambda_1*CE(y,y')+\lambda_2*Dice(y,y')
\end{equation}
where $\lambda_1$ and $\lambda_2$ are hyperparameters used to balance the two losses.

While training with cross-species data, the model's training process equates to handling a segmentation task with seven categories. Notably, the medulla category in the human dataset has significantly fewer images compared with other categories.  To address this class imbalance, we have replaced traditional Cross Entropy Loss with Focal Loss (FL) during the mixed training process. Additionally, we have implemented a specific weight $ w_i$ for each category $i$ within the Focal Loss framework, thereby boosting the model's ability to effectively handle imbalanced data by focusing more on the rare categories. Consequently, the overall loss function $L_t$ with hyperparameters $\lambda_3$ and $\lambda_4$ during joint training can be described as:

\begin{equation}
L_t=\lambda_3*w_i*FL(y,y')+\lambda_4*Dice(y,y')
\end{equation}

\subsection{Baseline}
\subsubsection{CNN-based Method}
To evaluate the effectiveness of cross-species homologous data on human kidney layer segmentation, we selected three CNN-based baseline models: U-Net \cite{ronneberger2015u}, tailored for medical image segmentation, alongside DeepLabv3+ \cite{chen2017rethinking} and Pyramid Scene Parsing Net (PSPNet)\cite{zhao2017pyramid}, both utilized for general image segmentation. We employed ResNet50 \cite{he2016deep} as the backbone for both U-Net and PSPNet to enhance feature extraction, while opting for MobileNetV2 \cite{sandler2018mobilenetv2} as the backbone for DeepLabv3+.

\subsubsection{Transformer-based Method}
 Transformer-based architectures have emerged as a prominent choice in the field since the appearance of the Vision Transformer (ViT) \cite{dosovitskiy2020image}. To assess the feasibility of our method across diverse network architectures, we selected two transformer-based models for evaluation: TransUNet \cite{chen2021transunet} and SwinUNet \cite{cao2022swin}. Specifically, we utilized ViT as the backbone for TransUNet, and Swin Transformer Tiny \cite{liu2021swin} for SwinUNet. Therefore, we can explore how cross-species homologous data performs on Transformer networks of varying parameter scales.

\begin{figure*}
\begin{center}
\includegraphics[width=\linewidth]{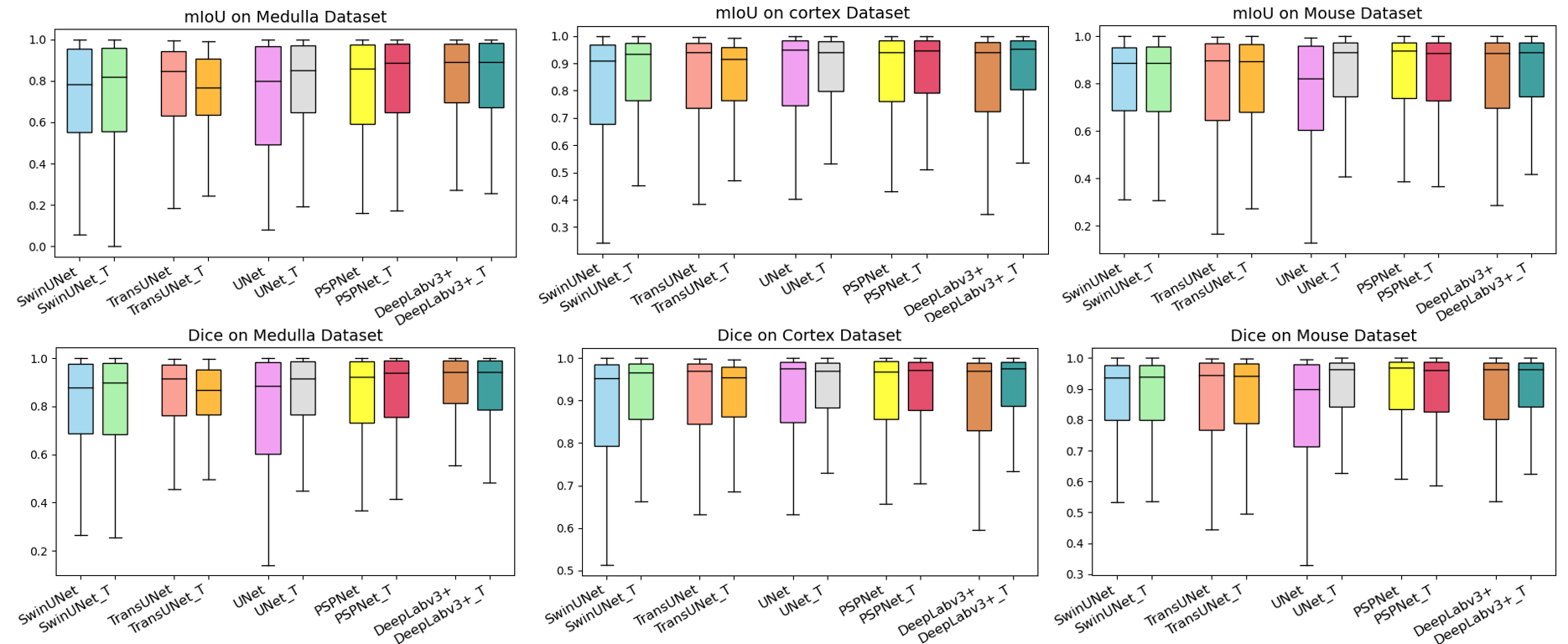}
\end{center}
\caption{The distribution of mIoU and Dice score across test datasets by different training manner. The collaborative training approach led to a more concentrated distribution of these metrics.}
\label{fig:Dice}
\end{figure*}

\section{EXPERIMENTS}
\label{sec:experiment}
\subsection{Data}
We collected a private PAS-stained pathology dataset comprising human and mouse kidney tissues sourced from both clinical and laboratory. For the human part, we obtained 27 WSIs from non-cancerous regions of nephrectomy samples. These images were scanned at 20X magnification and annotated by three professional clinicians using QuPath\cite{humphries2021qupath}, mapping out the contours of the medulla and cortex. Specific binary masks for both the cortex and medulla were derived from these contours. 

In the case of the mouse dataset, we acquired 30 renal pathology slides from male mice, including the wild-type mice, mice on the BKS background, and mice treated with an angiotensin receptor blocker. These images were scanned at 40X magnification and detailed annotated by clinicians in QuPath, marking the part of the inner medulla, inner stripe, outer stripe, and cortex.

Both datasets were initially divided into training and testing sets with a 7:3 ratio at the WSI level. Then, labels and whole slide images were segmented into 1024×1024 pixel patches. For the human dataset, this included 1,447 patches from the medulla and 2,714 from the cortex; the mouse dataset comprised 5,012 patches in total.

\subsection{Implementation}
The experiments were conducted on 4 NVIDIA Tesla V100 GPU cards.  We employed an SGD optimizer to configure with a momentum of 0.9 and a weight decay parameter set at $10^{-4}$. The initial learning rate, denoted as $lr_0$, was established at $1 \times 10^{-3}$. A cosine decay strategy is applied to reduce the learning rate to 0.01 of the initial rate during the training process. We have fine-tuned the hyperparameters $\lambda_1$, $\lambda_2$, $\lambda_3$, and $\lambda_4$ in our hybrid loss function to values of 0.5, 0.75, 0.75, and 1.0, respectively. Additionally, within the Focal loss component, we adjusted the weights for the human cortex and medulla to 1.25 and 2.0, respectively, while setting the weights for other categories to 1.

\begin{table}
\centering
\caption{The quantitative performance of models on various datasets with different training methods. Compared with training independently, there is an average improvement of 1.77\% and 1.24\% in mIoU, along with 1.76\% and 0.89\% in Dice score on the human cortex and medulla datasets respectively.}
    \begin{tabular}{cccccccc}
    \hline
        Model & Dataset & Separate & Together & mIoU & Dsc & std IoU & std Dsc  \\ \hline
         & mouse & \checkmark & ~ & 0.7644 & 0.8282 & 0.0438 & 0.0322  \\ 
        ~ & ~ & ~ & \checkmark & 0.8594 & 0.8927 & 0.0316 & 0.0187  \\ 
        UNet\cite{ronneberger2015u} & medulla & \checkmark & ~ & 0.7163 & 0.7874 & 0.0704 & 0.0495  \\ 
         & ~ & ~ & \checkmark & 0.7860 & 0.8477 & 0.0452 & 0.0310  \\ \
        ~ & cortex & \checkmark & ~ & 0.8405 & 0.8851 & 0.0461 & 0.0314  \\ 
        ~ & ~ & ~ & \checkmark & 0.8458 & 0.9034 & 0.0279 & 0.0212  \\ \hline  \hline
         & mouse & \checkmark & ~ & 0.8495 & 0.8958 & 0.0290 & 0.0186  \\ 
        ~ & ~ & ~ & ~\checkmark& 0.8412 & 0.8891 & 0.0296 & 0.0196  \\ 
        PSPNet\cite{zhao2017pyramid} & medulla & \checkmark & ~ & 0.7767 & 0.8375 & 0.0503 & 0.0353  \\
        ~ & ~ & ~ & \checkmark & 0.7972 & 0.8505 & 0.0457 & 0.0328  \\ 
        ~ & cortex &\checkmark & ~ & 0.8406 & 0.8853 & 0.0433 & 0.0303  \\ 
        ~ & ~ & ~ &\checkmark & 0.8551 & 0.8962 & 0.0361 & 0.0255  \\ \hline \hline
         & mouse & \checkmark & ~ & 0.8308 & 0.8810 & 0.0336 & 0.0217  \\
        ~ & ~ & ~ & \checkmark & 0.8509 & 0.8973 & 0.0255 & 0.0165  \\ 
        DeepLabv3+\cite{chen2017rethinking} & medulla & \checkmark & ~ & 0.8130 & 0.8701 & 0.0391 & 0.0255  \\ 
        ~ & ~ & ~ & \checkmark & 0.8042 & 0.8580 & 0.0458 & 0.0323  \\ 
        ~ & cortex & \checkmark & ~ & 0.8325 & 0.8780 & 0.0436 & 0.0314  \\ 
        ~ & ~ & ~ & \checkmark& 0.8667 & 0.9062 & 0.0331 & 0.0226  \\  \hline \hline
         & mouse & \checkmark & ~ & 0.8022 & 0.8566 & 0.0405 & 0.0288  \\
                ~ & ~ & ~ & \checkmark& 0.8099 & 0.8629 & 0.0319 & 0.0256  \\
        TransUNet\cite{chen2021transunet} & medulla & \checkmark & ~ & 0.7785 & 0.8503 & 0.0361 & 0.0231  \\ 
        ~ & ~ & ~ & \checkmark & 0.7454 & 0.8259 & 0.0359 & 0.0256  \\
        ~ & cortex & \checkmark & ~ & 0.8349 & 0.8825 & 0.0382 & 0.0268  \\ 
        ~ & ~ & ~ & \checkmark & 0.8330 & 0.8845 & 0.0351 & 0.0237  \\ \hline \hline
         & mouse & \checkmark & ~ & 0.8105 & 0.8701 & 0.0306 & 0.0199  \\
        ~ & ~ & ~ & \checkmark & 0.8124 & 0.8707 & 0.0304 & 0.02  \\ 
        SwinUNet\cite{cao2022swin} & medulla & \checkmark & ~ & 0.7273 & 0.7998 & 0.0611 & 0.0443  \\
        ~ & ~ & ~ & \checkmark & 0.7410 & 0.8076 & 0.0614 & 0.0470  \\ 
        ~ & cortex & \checkmark & ~ & 0.8016 & 0.8563 & 0.0527 & 0.0383  \\ 
        ~ & ~ & ~ & \checkmark & 0.8379 & 0.8849 & 0.0419 & 0.0291  \\ \hline
        
\label{metrics}
\end{tabular}
\end{table}

\section{RESULTS}
\label{sec:result}  

The evaluation metrics used were Mean Intersection over Union (mIoU) and Dice score, with higher values indicating better semantic segmentation performance. The quantitative performance of models is presented in Table. \ref{metrics}. By integrating homologous mouse kidney data with human datasets for combined training, all models demonstrated an average improvement of 1.77\% and 1.24\% in mIoU, along with 1.76\% and 0.89\% in Dice score on the human cortex and human medulla datasets respectively. The distribution of mIoU and Dice score across testsets before and after joint training is depicted in the boxplot in Figure. \ref{fig:Dice}. The collaborative training approach led to a more concentrated distribution of these metrics. Further, observing changes in the standard deviations of mIoU and Dice score in Table. \ref{metrics}, the incorporation of cross-species homologous data markedly enhanced the segmentation performance of both CNN-based and Transformer-based models in human kidney layer segmentation, and also significantly boosted the generalization capabilities of models.

\begin{figure*}
\begin{center}
\includegraphics[width=0.9\linewidth]{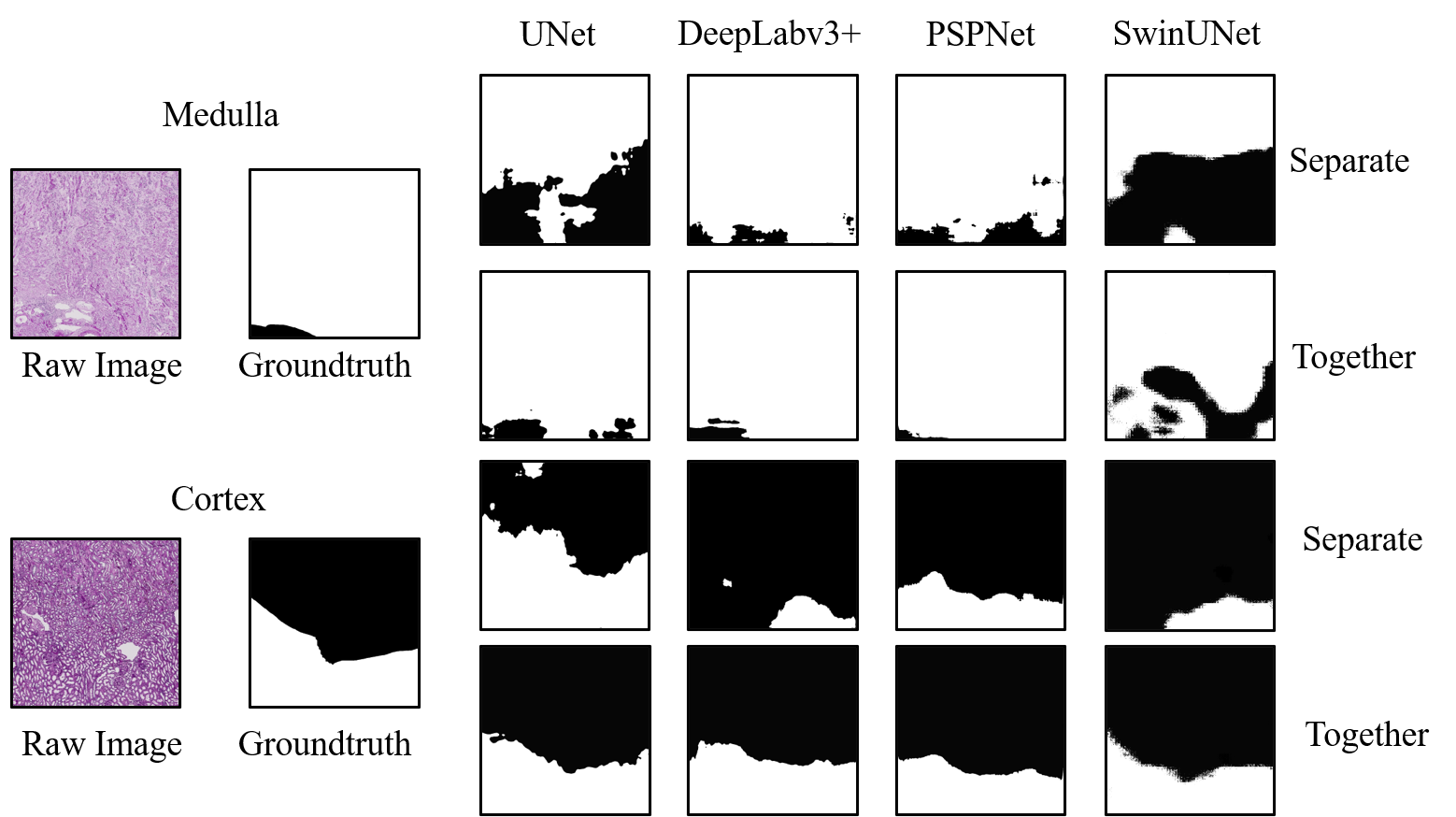}
\end{center}
\caption{The qualitative outcomes of models on various datasets. By utilizing external homologous data, the models have become better at perceiving edge textures, thus performing better in more precise localization and identification of kidney layer boundaries.}
\label{fig:seg}
\end{figure*}

Figure. \ref{fig:seg} displays the qualitative outcomes of models on various datasets. Incorporating external homologous data has improved the models' capacity for edge texture perception, which has led to more accurate localization and identification of kidney layer boundaries, thereby enhancing overall model performance.

\section{CONCLUSION}
\label{sec:conclusion}  
In this study, we demonstrated that cross-species homologous data can enhance layer segmentation in the human kidney. Mouse kidney data, with its high structural and staining similarity, significantly improved the semantic segmentation performance of both CNN and Transformer-based models on the human medulla and cortex layers and also boosted the models’ generalization abilities. This finding opens up new avenues for medical image analysis by providing an alternative source of data. Integrating homologous data from different species into studies that focus on a single species can extend the pool of low-noise, trainable data. This reduces the limitations imposed by privacy concerns in medical data on human disease research and allows for a better understanding and simulation of complex biological processes when clinical samples are limited.

\acknowledgments 
This research was supported by NIH R01DK135597(Huo), DoD HT9425-23-1-0003(HCY), NIH NIDDK DK56942 (ABF). This work was also supported by Vanderbilt Seed Success Grant, Vanderbilt Discovery Grant, and VISE Seed Grant. This project was supported by The Leona M. and Harry B. Helmsley Charitable Trust grant G-1903-03793 and G-2103-05128. This research was also supported by NIH grants R01EB033385, R01DK132338, REB017230, R01MH125931, and NSF 2040462. We extend gratitude to NVIDIA for their support by means of the NVIDIA hardware grant.

\bibliography{report} 
\bibliographystyle{spiebib} 

\end{document}